\documentclass[
reprint, 
superscriptaddress,
 amsmath,amssymb,
 pre,
]{revtex4-1}

\usepackage{graphicx}
\usepackage{dcolumn}
\usepackage{bm}
\usepackage{todonotes}

\usepackage{color}

\newcommand{\diff}[2]{\frac{d#1}{d#2}}

\newcommand{\abs}[1]{\left|#1\right|}

\newcommand{\be}{\bm{e}}

\newcommand{\br}{\bm{r}}

\newcommand{\ave}[1]{\left\langle #1 \right\rangle}

\newcommand{\FF}{\mathcal{F}}

\newcommand{\DD}{\mathcal{D}}
\newcommand{\PP}{\mathcal{P}}

\usepackage{amsmath}	
\begin{document}

\preprint{APS/123-Qed} \title{Non-Affine Displacements Below Jamming
under Athermal Quasi-Static Compression}


\author{Harukuni Ikeda}
 \email{hikeda@g.ecc.u-tokyo.ac.jp}
\affiliation{Graduate School of Arts and Sciences, The University of
Tokyo 153-8902, Japan}
 \author{Koji Hukushima}
 \email{k-hukushima@g.ecc.u-tokyo.ac.jp}
\affiliation{Graduate School of Arts and Sciences, The University of
Tokyo 153-8902, Japan} \affiliation{Komaba Institute for Science, The
University of Tokyo, 3-8-1 Komaba, Meguro, Tokyo 153-8902, Japan}


\date{\today}
	     
\begin{abstract}  
Critical properties of frictionless spherical particles below jamming
are studied using extensive numerical simulations, paying particular
attention to the non-affine part of the displacements during the
athermal quasi-static compression. It is shown that the squared norm of
the non-affine displacement exhibits a power-law divergence toward the
jamming transition point. A possible connection between this critical
exponent and that of the shear viscosity is discussed. The participation
ratio of the displacements vanishes in the thermodynamic limit at the
transition point, meaning that the non-affine displacements are
localized marginally with a fractal dimension. Furthermore, the
distribution of the displacement is shown to have a power-law tail, the
exponent of which is related to the fractal dimension.
\end{abstract}

\maketitle

\section{Introduction}

Considering the process of increasing the density of 
particle system at zero temperature, if the density is low enough, the
particles do not overlap. At a certain density, the particles begin to
come into contact, and as a result, the system suddenly gains finite
energy, mechanical pressure, and stiffness without any apparent
structural changes~\cite{van2009}.  This phenomenon called jamming has
been actively studied in recent years, and the onset is defined as the
jamming transition point $\varphi_J$. The jamming transition is
ubiquitously observed for very diverse athermal systems such as metallic
bolls~\cite{bernal1960packing},
forms~\cite{durian1995,katgert2010jamming}, colloids~\cite{zhang2009},
polymers~\cite{karayiannis2009}, candies~\cite{donev2004},
dices~\cite{jaoshvili2010}, biological tissues~\cite{bi2015}, growing
microbes~\cite{delarue2016self}, and some neural
networks~\cite{franz2016simplest,franz2019jamming}.

A famous and popular numerical protocol to generate a jamming
configuration is the athermal quasi-static compression (AQC), which
combines the affine transformation with successive energy
minimization~\cite{ohern2003}. An advantage of this protocol is that one
can unambiguously define the jamming transition point $\varphi_J$ as the
packing fraction $\varphi$ at which the energy after the minimization
has a non-zero finite value. With the AQC, extensive work has been done
for frictionless, spherical, and purely repulsive particles above
jamming $\varphi>\varphi_J$. Systematic numerical studies including
finite size scaling analyses determine the precise values of the
critical exponents in two and three spatial
dimensions~\cite{ohern2003,goodrich2012,charbonneau2015jamming}, and
quasi-one dimension~\cite{hikeda2020jamming}. The numerical results of
the critical exponents well agree with the mean-field predictions in two
and three dimensions~\cite{wyart2005,charbonneau2014fractal}.

Contrarily and somewhat surprisingly, the critical properties of the
jamming transition below $\varphi_J$ during the AQC have not yet been
fully investigated even for frictionless spherical particles.  One of
the reasons is that the quantities showing the criticality above
$\varphi_J$, such as the mechanical pressure, energy, and bulk/shear
modulus, are trivially zero below $\varphi_J$, and other appropriate
quantities are not necessarily clear below
$\varphi_J$~\cite{ohern2003}. The criticality below $\varphi_J$ has been
mainly investigated by adding thermal
fluctuation~\cite{ikeda2013dynamic,charbonneau2014fractal}, introducing
a moving tracer~\cite{drocco2005}, considering self-propelled
particles~\cite{liao2018criticality}, or quenching from random initial
configurations~\cite{ikeda2020univ,nishikawa2020relaxation}. In
particular, extensive work has been conducted on shear-driven
systems~\cite{marty2005subdiffusion,olsson2007critical,
hatano2008scaling,heussinger2009,lerner2012unified,kawasaki2015,olsson2019dimension}.
However, it would be more desirable if one can directly extract the
criticality from the configurations during the AQC. A promising study in
this direction has been done by Shen \textit{et al.}~\cite{shen2012}.
They observed a rapid increase of several physical quantities, such as
the displacements of the particle positions, just below
$\varphi_J$. However, the critical exponent below $\varphi_J$ under the
AQC has not been calculated yet.

In this work, we characterize the criticality below $\varphi_J$ during
the AQC by investigating the statistical properties of the non-affine
displacements for frictionless spherical particles in three
dimensions. We show that the mean square of the non-affine
displacements diverges toward $\varphi_J$ with the critical exponent
very close to that of the shear viscosity. By observing the
participation ratio, it is shown that the displacements become more
localized as the system approaches $\varphi_J$. Furthermore, by using
the finite size scaling of the participation ratio, we calculate the
fractal dimension of the displacements at $\varphi_J$.  Finally, we 
show that 
the distribution of the non-affine displacements 
has a power-law tail at $\varphi_J$, and prove that
there is a scaling relation between the power-law tail and fractal
dimension.

\section{Model}

We consider $N$ frictionless spherical particles in three
dimensions. The interaction potential between $N$ particles is given as
\begin{align}
 V = \sum_{i<j}^{1,N}\frac{h_{ij}^2}{2}\theta(-h_{ij}), 
 h_{ij} = \abs{\br_i-\br_j}-R_i-R_j,\label{071430_8Aug20}
\end{align}
where 
$\br_i=\{x_i,y_i,z_i\}$ and $R_i$
respectively denote the position and radius of the $i$-th particle. 
The particles are confined 
in a cubic box $\br_i\in [0,L]^3$ with the
periodic boundary conditions in all directions. To avoid
crystallization, we use a binary mixture consisting of ${N_{\rm
S}=N/2}$ small particles and ${N_{L}=N/2}$ large particles. The radii of
large and small particles are $R_{\rm S}=0.5$ and $R_{\rm L}=0.7$,
respectively. With those notations, the volume fraction $\varphi$ is
written as
\begin{align}
\varphi = \frac{2\pi N(R_{\rm S}^3 + R_{\rm L}^3)}{3L^3}.
\end{align}
The settings of this model are standard ones that have been studied in
the context of the jamming transition~\cite{ohern2003}.

\section{Numerical simulations}
Here we describe the AQC originally proposed by O'Hern \textit{et
al.}~\cite{ohern2003}.  Starting from a random initial configuration at
a small packing fraction, for example $\varphi=0.1$, compression and
energy minimization are performed successively in sequence.  For each
step of the compression, the packing fraction is slightly increased as
$\varphi\to \varphi+\varepsilon$ with $\varepsilon=10^{-5}$, by
performing the affine transformation $\br_i\to \br_i
L(\varphi+\varepsilon)/L(\varphi)$, where $L(\varphi)= \left[2\pi
N(R_{\rm S}^3 + R_{\rm L}^3)/3\varphi\right]^{1/3}$. Then, the energy is
minimized by using the FIRE algorithm, for details see
Ref.~\cite{bitzek2006}, until the energy or squared force becomes
sufficiently small: $V_N/N<10^{-16}$ or $\sum_{i=1}^N(\nabla_i
V_N)^2/N<10^{-25}$. The procedure is repeated up to the jamming
transition point $\varphi_J$ at which $V_N/N>10^{-16}$ after the
minimization~\cite{ohern2003}.

We perform the numerical simulations for various system sizes $N=256$,
$512$, $1024$, $2048$, and $4096$. To improve the statistics, we average
over $1000$ samples for $N=256$ and $100$ samples for the
other system sizes. We confirmed that the results do not depend on
$\varepsilon$, see Appendix~\ref{053220_24Dec20}.

\section{Mean squared displacement}
\label{110649_26Dec20}

When the system is compressed from $\varphi$ to $\varphi + \varepsilon$,
the displacement of the $i$-th particle can be written as
\begin{align}
\br_i(\varphi+\varepsilon) -\br_i(\varphi)= \delta\br_i^{\rm A} +
\delta\br_i^{\rm NA}
\end{align}
where $\delta\br_i^{\rm
A}=\left[L(\varphi+\varepsilon)/L(\varphi)-1\right]\br_i(\varphi)$ and
$\delta\br_i^{\rm NA}$ respectively denote the affine and non-affine
parts of the displacement. In this work, we only focus on the
non-trivial part of the displacement $\delta\br_i^{\rm NA}$.
\begin{figure}[t]
\begin{center}
\includegraphics[width=8cm]{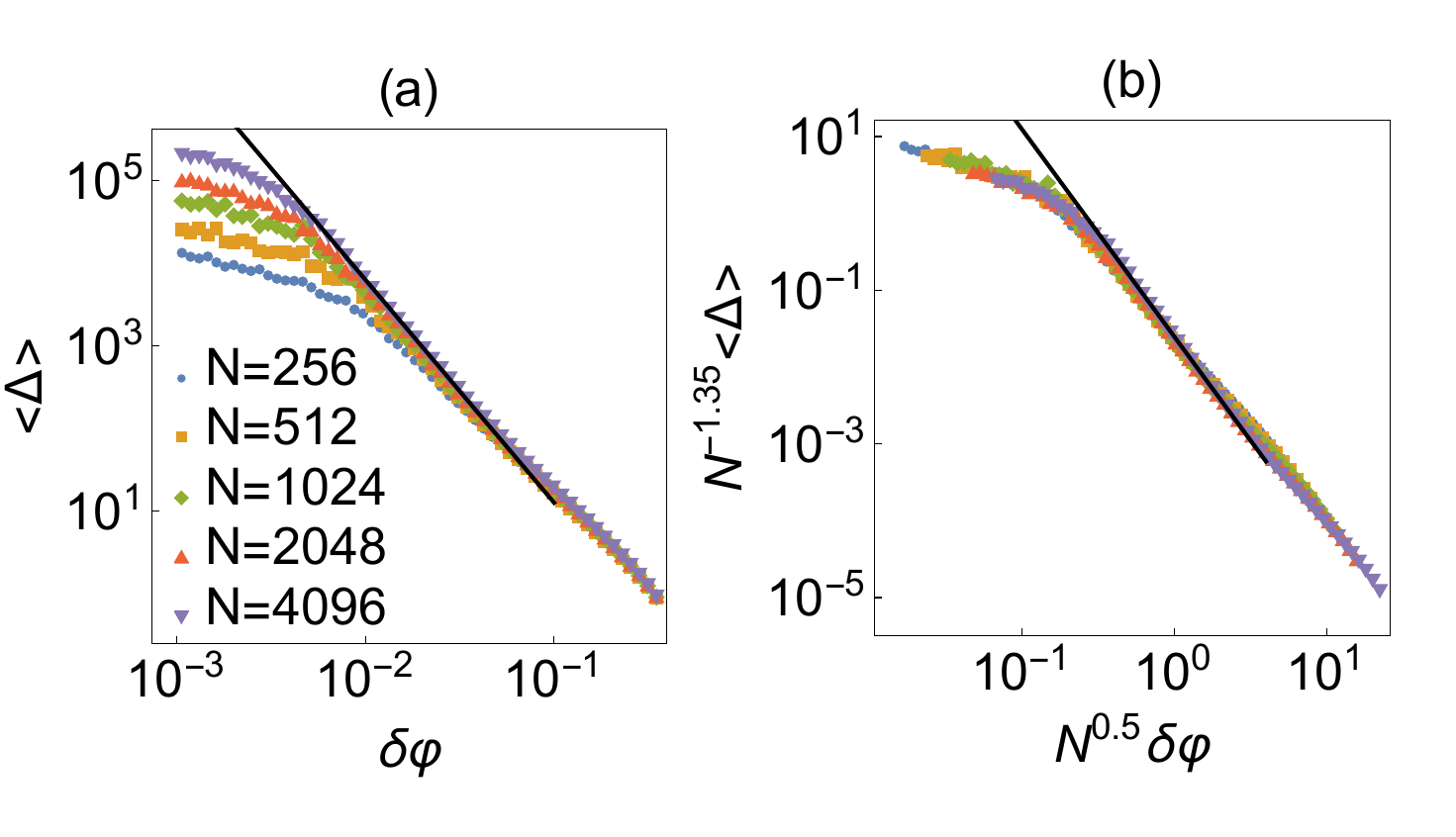} \caption{(a) Mean squared
non-affine displacement $\ave{\Delta}$. Markers denote numerical
results, while the solid line denotes the power law ${\Delta\propto
\delta\varphi^{-2.7}}$.  (b) Scaling plot of the same data as in (a). }
\label{185522_22Jul20}
\end{center}
\end{figure}

To characterize the criticality of the non-affine displacements, we
first observe the mean squared displacement
\begin{align}
 \ave{\Delta} = \frac{1}{N}\sum_{i=1}^N \Delta_i,\label{102414_26Dec20}
\end{align}
with
\begin{align}
\Delta_i = \frac{\left(\delta\br_i^{\rm NA}\right)^2}{3\delta l^2},
\end{align}
where $\delta l=L(\varphi+\varepsilon)/L(\varphi)-1$ accounts for the
change of the linear distance $L$ of the system. In
Fig.~\ref{185522_22Jul20}~(a), we show $\ave{\Delta}$ as a function of
$\delta\varphi = \varphi_J-\varphi$. For large $N$ and intermediate
$\delta\varphi$, $\ave{\Delta}$ shows the power law
\begin{align}
\ave{\Delta}\propto
\delta\varphi^{-\beta}.
\end{align}
The power-law fitting of the data for $N=4096$ in the range
$\delta\varphi \in (0.01,0.1)$ leads to
\begin{align}
\beta = 2.7,\label{055344_24Dec20}
\end{align}
see the solid line in Fig.~\ref{185522_22Jul20}~(a). Interestingly,
Eq.~(\ref{055344_24Dec20}) is close to the theoretical prediction for
the critical exponent of the shear viscosity
$\beta'=2.8$~\cite{PhysRevE.91.062206}. We shall discuss a possible
connection between $\ave{\Delta}$ and shear viscosity later in this
paper.

To further investigate the scaling of $\ave{\Delta}$, we perform a
finite-size scaling analysis assuming the following scaling function:
\begin{align}
 \ave{\Delta} = N^\alpha \DD(N^{\alpha/\beta} \delta\varphi),\label{142242_24Jul20}
\end{align}
where $\DD(x)=O(1)$ for $x\ll 1$, and $\DD(x)\sim x^{-\beta}$ for $x\gg
1$. As shown in Fig.~\ref{185522_22Jul20}~(b), a good scaling collapse
is obtained with $\alpha=1.35$. This result implies that the number of
the correlated particles diverges as $N_{\rm cor}\sim
\delta\varphi^{-\beta/\alpha}\sim \delta\varphi^{-2}$, and the
correlation length $L_{\rm cor}\sim N_{\rm cor}^{1/3}\sim
\delta\varphi^{-\nu}$ with $\nu=2/3\approx 0.67$ under the assumption
that the correlated volume is compact. This is close to a previous
result $\nu=0.71$ obtained by the finite-size scaling analysis of the
transition point~\cite{ohern2003}.

\section{Participation ratio}

To see the spatial structure of the displacements, we observe the
normalized vector~\cite{yamamoto1998}:
\begin{align}
 \be_i = \frac{\delta \br_i^{\rm NA}}{\sqrt{\sum_{i=1}^N \left(\delta \br_i^{\rm NA}\right)^2}},
\end{align}
which satisfies $\sum_{i=1}^N \be_i\cdot\be_i=1$.
\begin{figure*}[t]
\begin{center}
\includegraphics[width=18cm]{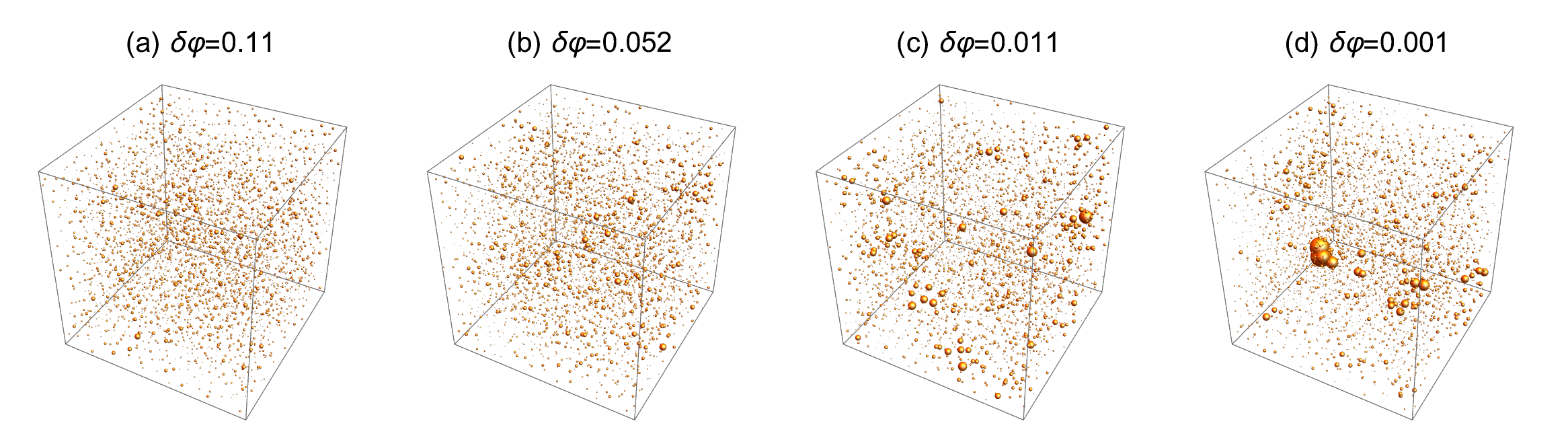} \caption{ Spatial distribution
of non-affine displacements for $N=4092$ particles. Diameters of spheres
represent the amplitude of the (normalized) non-affine displacements
$\abs{\be_i}$.}  \label{095853_27Jul20}
\end{center}
\end{figure*}
In Fig.~\ref{095853_27Jul20}, we visualize $\be_i$ by drawing spheres
such that their radii are proportional to $\abs{\be_i}$. Far from
jamming, the spatial distribution of $\be_i$ is homogeneous and
featureless, see Fig.~\ref{095853_27Jul20}~(a). On the contrary, near
jamming, a few particles have very large displacements, and thus the
displacement is spatially heterogeneous and localized, see
Fig.~\ref{095853_27Jul20}~(d).
\begin{figure}[t]
\begin{center}
\includegraphics[width=8cm]{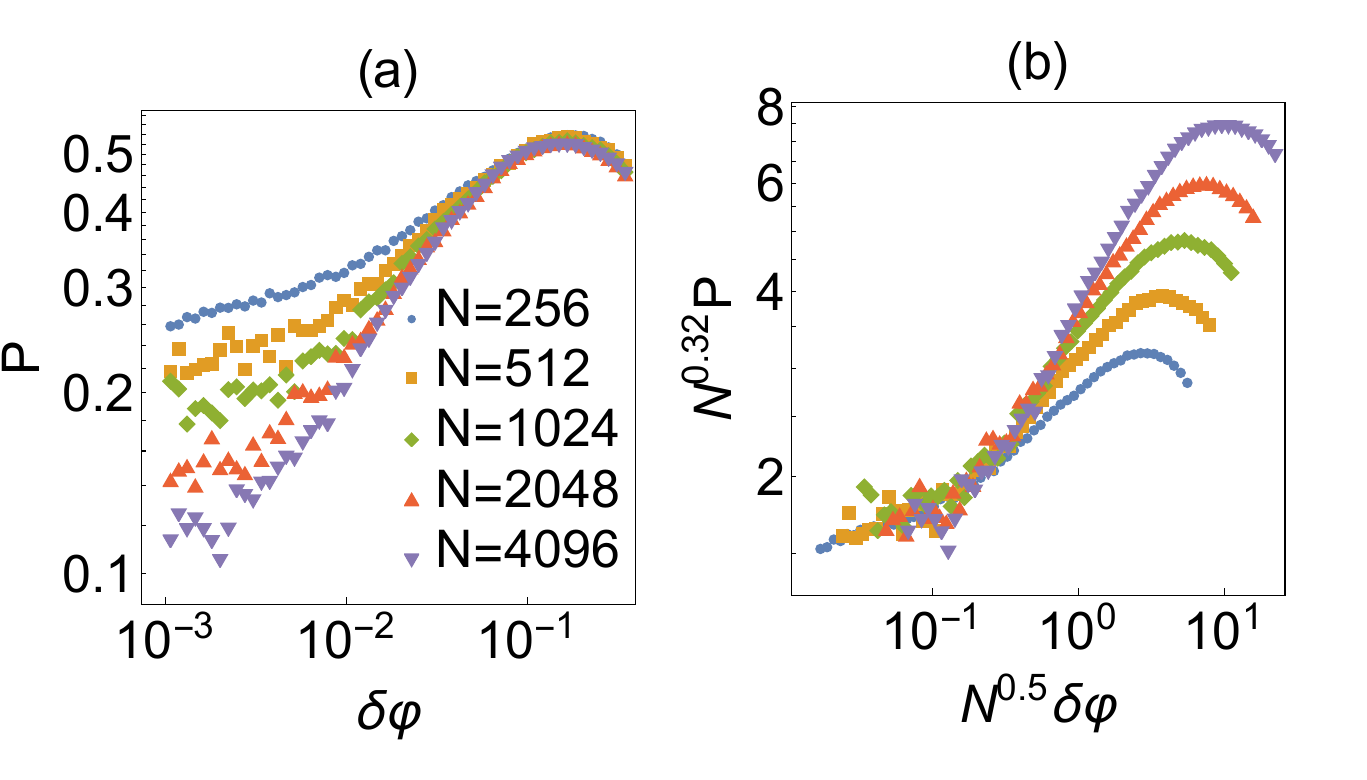} \caption{(a) Participation ratio
$P$. Markers denote numerical results. (b) Scaling plot of the
same data.}  \label{032747_23Jul20}
\end{center}
\end{figure} 
We quantify the degree of the localization by using the participation
ratio:
\begin{align}
 P = \frac{1}{N}
 \frac{\left(\sum_{i=1}^N \be_i\cdot\be_i\right)^2}{\sum_{i=1}^N \left(\be_i\cdot\be_i\right)^2},\label{171919_19Aug20}
\end{align}
which (or inverse of which) is widely used in the study of condensed
matter physics~\cite{kramer1993localization}, including amorphous
solids~\cite{lerner2016,mizuno2017continuum}. If $\be_i$ is spatially
localized to a single particle, say ${\be_i\cdot\be_i = \delta_{i1}}$,
the participation ratio $P$ is proportional to $N^{-1}$.  On the
contrary, if $\be_i$ is extended such that ${\be_i\cdot\be_i\sim
N^{-1}}$ for all $i$, $P$ is constant independent of $N$.  In
Fig.~\ref{032747_23Jul20}~(a), we show the $\delta\varphi$ dependence of
$P$. One can see that $P$ decreases with approaching $\varphi_J$ and
increasing $N$. To investigate the $N$ dependence of $P$, we use the
following scaling form:
\begin{align}
P = N^{-\gamma}\PP(N^{\alpha/\beta}\delta\varphi),\label{173356_19Aug20}
\end{align}
assuming the same correlated volume as in Eq.~(\ref{142242_24Jul20}). We
find a good collapse with
\begin{align}
\gamma=0.32\label{081726_24Dec20}
\end{align}
near the jamming point, see Fig.~\ref{032747_23Jul20}~(b). At
$\varphi_J$, $P$ vanishes in the thermodynamic limit as $P\sim
N^{-\gamma}$. This exponent relates to the fractal
dimension $d_f$ of $\be_i$, namely, if $\be_i\cdot\be_i \sim L^{-d_f}$
for $i=1,\dots, L^{d_f}$, this yields that $P\sim
N^{-1+d_f/d}$~\cite{kramer1993localization}, leading to
\begin{align}
d_f=3(1-\gamma) = 2.04.\label{005239_8Aug20}
\end{align}
Therefore, $\be_i$ has a more compact structure than the bulk $d=3$. A
mean-field theory of the jamming transition predicts that the correlated
volume $v_{\rm col}$ and correlation length $l_{\rm cor}$ have the
following relation $v_{\rm col}\sim l_{\rm
cor}^2$~\cite{yan2016variational,during2016}. This may imply that the
fractal dimension is $d_{f}=2$, which is close to our estimation
Eq.~(\ref{005239_8Aug20}).

Interestingly, the similar spatially heterogeneous structures are
observed for super-cooled liquids near the glass transition
point~\cite{ediger2000spatially}. For the studies of the glass
transition, the degree of spatial heterogeneity is characterized by the
so-called non-gaussian parameter~\cite{rahman1964,kob1996test}. In our
setting, an analogous quantity may be written as
\begin{align}
 \alpha_2 = \frac{3\ave{(\delta\br^{\rm NA})^4}}{5\ave{(\delta\br^{\rm NA})^2}^2}-1 
= \frac{3}{5P}-1. 
\end{align}
If the displacements follow the featureless gaussian distribution, one
obtains $\alpha_2=0$. For the supercooled liquids, $\alpha_2$ of the
displacements rapidly increases on decreasing the
temperature~\cite{kob1996test,weeks2000three}. Similarly, $\alpha_2$ of
our model increases on approaching $\varphi_J$, because $P\to 0$ and
$\alpha_2\propto P^{-1}$. Furthermore, an experimental study for the
supercooled colloidal suspensions, which approximately behave as hard
spheres~\cite{pusey1986phase,van1994} and may have the same interaction
as our model below jamming, showed that the dynamically correlated
regions form a compact cluster of the fractal dimension $d_f=1.9\pm
0.4$~\cite{weeks2000three}, reasonably close to our result of
Eq.~(\ref{005239_8Aug20}).  Also, the inhomogeneous mode-coupling
theory, which is a mean-field theory of the glass transition, predicts
$d_f=2$ in the early stage of the relaxation
process~\cite{inhomogeneous2006}. Those results suggest the existence of
the underlying universality between the dynamics of the athermal system
near $\varphi_J$ and thermal systems near the glass transition
point~\cite{marty2005,chaudhuri2007}.

\section{Distribution function}

\begin{figure}[t]
\begin{center}
\includegraphics[width=8cm]{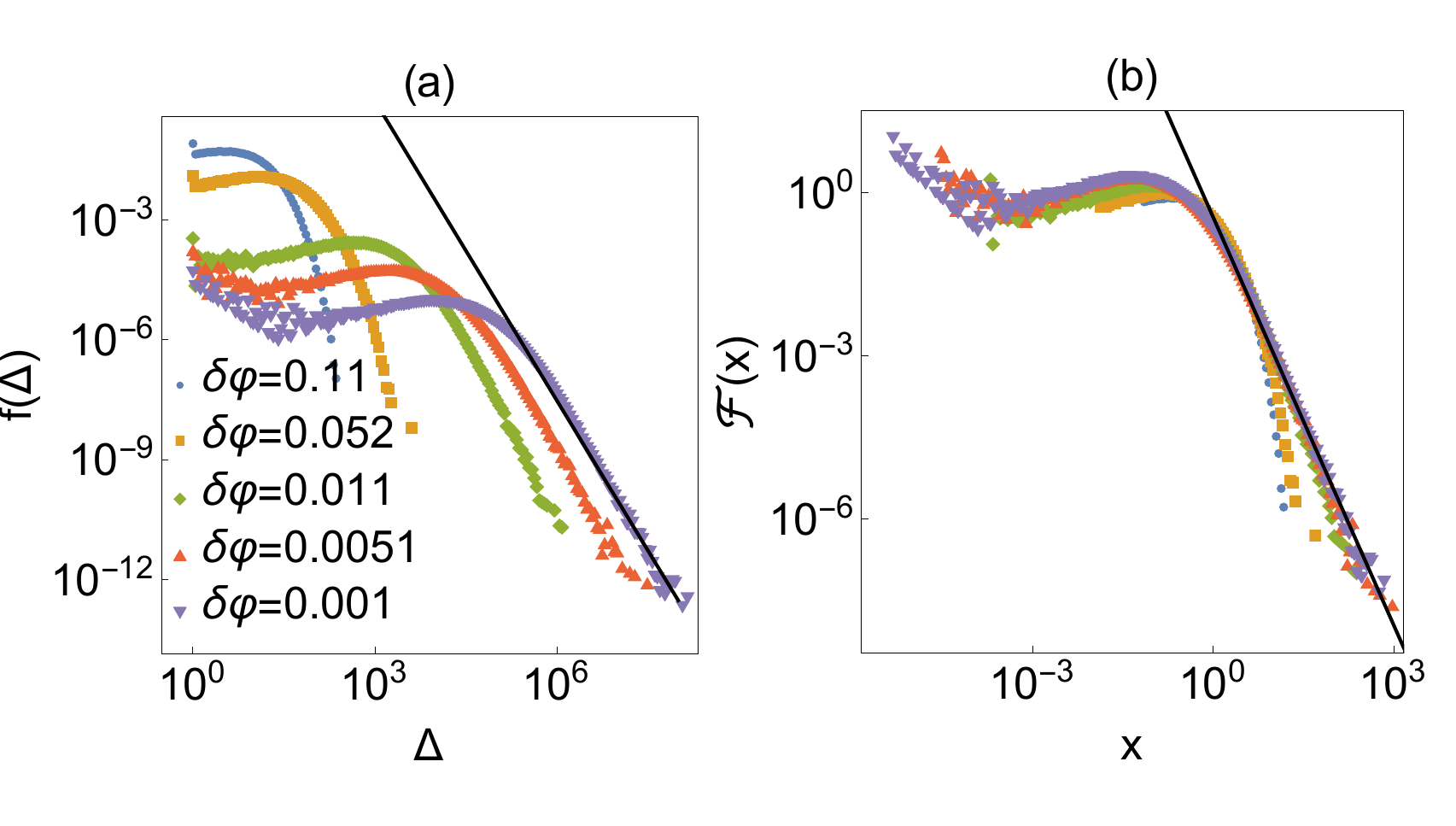} \caption{(a) Distribution of
$\Delta$ for $N=4096$. Markers are numerical results, while the solid
line denotes the power law ${f(\Delta)\sim \Delta^{-2.5}}$. (b)
Distribution of $x=\Delta/\ave{\Delta}$ for the same data. The solid
line denotes ${\FF(x)\sim x^{-2.5}}$. } \label{195640_23Jul20}
\end{center}
\end{figure}

Here, we discuss the behavior of $P$ from the perspective of the
distribution of $\Delta$. First, we note that $P$ can be written as
\begin{align}
 P = \frac{\left[\int_0^\infty d\Delta f(\Delta)\Delta\right]^2}{\int_0^\infty d\Delta f(\Delta)\Delta^2},\label{210610_16Aug20}
\end{align}
where the distribution $f(\Delta)$ of $\Delta$ is defined as 
\begin{align}
 f(\Delta) = \frac{1}{N}\sum_{i=1}^N \delta(\Delta-\Delta_i).
\end{align}
Fig.~\ref{195640_23Jul20}~(a) presents $f(\Delta)$ for several
$\delta\varphi$. $f(\Delta)$ has a broader distribution for smaller
$\delta\varphi$. For the later convenience, we define a scaled variable
$x=\Delta/\ave{\Delta}$, and distribution function
\begin{align} 
\FF(x) = f(\Delta)\diff{\Delta}{x} = \ave{\Delta}f(\ave{\Delta}x).
\label{010828_24Jul20}
\end{align}
By definition, $\int dx \FF(x) = \int dx \FF(x)x = 1$. 
As shown in Fig.~\ref{195640_23Jul20}~(b),
with decreasing $\delta\varphi$, 
$\FF(x)$ develops the power-law tail
\begin{align}
 \lim_{\delta\varphi\to 0}\FF(x)\sim x^{-\zeta}\ {\rm for}\ x\gg 1.
\end{align}
A similar fat-tail was previously reported for the velocity distribution
of sheared driven systems in the quasi-static limit near
$\varphi_J$~\cite{tighe2010,andreotti2012,olsson2016}.  Now, we show
that the exponent $\zeta$ relates to $\gamma$ in
Eq.~(\ref{173356_19Aug20}). Using Eq.~(\ref{210610_16Aug20}) and
(\ref{010828_24Jul20}), we get
\begin{align}
 P = \left(\int_0^\infty dx x^2 \FF(x)\right)^{-1}.
\end{align}
If $\zeta < 3$, the denominator diverges, leading to $P=0$ at
$\varphi_J$. For finite $N$, however, the divergence does not occur as
the power law of $\FF(x)$ is truncated at finite $x_{\rm max}$.  Using
the extreme value statistics, we can calculate $x_{\rm max}$ as
\begin{align}
 \int_{x_{\rm max}}^{\infty}\FF(x)dx \sim \frac{1}{N}\to x_{\rm max} \sim N^{\frac{1}{\zeta-1}}.
\end{align}
Then, $P$ for finite $N$ is expressed as  
\begin{align}
 P\sim \left(\int_0^{x_{\rm max}} dx x^2 \FF(x)\right)^{-1}\sim N^{-\frac{3-\zeta}{\zeta-1}}.
\end{align}
Comparing this with Eq.~(\ref{173356_19Aug20}) for $\delta\varphi=0$, we
finally get
\begin{align}
 \zeta = \frac{3+\gamma}{1+\gamma} = 2.5.\label{073355_24Dec20}
\end{align}
This is consistent with the assumption $\zeta<3$ and well agrees with
the numerical result, see Fig.~\ref{195640_23Jul20}.

\section{Summary and discussions}

In summary, we investigated the statistical properties of the non-affine
displacements under the AQC below the jamming transition point. We showed that the
mean square of the non-affine displacement diverges toward the jamming
transition point. At the jamming transition point, the distribution of
the non-affine displacements has a power-law tail, the exponent of which
relates to the fractal dimension.

An interesting question is how the present work relates to the previous
works for the shear driven systems. As the shear viscosity $\eta$
diverges with the same critical exponent as the bulk viscosity $\eta_p$
near $\varphi_J$, namely, $\eta\sim\eta_p\sim
\delta\varphi^{-\beta'}$~\cite{olsson2012,vergberg2014universal}, we
consider a system compressed with a finite compression rate
$\dot{l}=\dot{L}/L$, instead of the shear driven system. The work done
by the imposed compression per time is $ p\dot{l}L^3=
\eta_p\dot{l}^2L^3$, where $p$ denotes the pressure, and $\eta_p =
p/\dot{l}$ denotes the bulk viscosity. In the quasi-static limit
$\dot{l}\to 0$, this should be balanced with the dissipation
$\sum_{i=1}^N\left(\dot{\br}_i^{\rm NA}\right)^2$, leading
to~\cite{lerner2012unified,katgert2013jamming}
\begin{align}
 \eta_p \sim \frac{1}{L^3}\sum_{i=1}^N\left(\frac{\dot{\br}_i^{\rm NA}}{\dot{l}}\right)^2
 \sim \frac{1}{L^3}\sum_{i=1}^N\left(\frac{\delta\br_i^{\rm NA}}{\delta l}\right)^2 \sim \ave{\Delta}.
\end{align}
A mean-field theory of sheared suspensions predicts $\eta\sim
\delta\varphi^{-\beta'}$ with $\beta'=2.8$~\cite{PhysRevE.91.062206},
which is close to our result ${\ave{\Delta}\sim \delta\varphi^{-\beta}}$
with $\beta=2.7$ and thus supports the above conjecture.  However, the
numerical result of $\beta'$ varies widely from one literature to another.
For instance, Ref.~\cite{kawasaki2015} reported $\beta'=2.55$, while
Ref.~\cite{olsson2019dimension} reported $\beta'=3.82$. Further study is
necessary to elucidate this point.

From a practical point of view, the AQC has several advantages over
other dynamical methods such as implying shear to characterize the
criticality below jamming. First of all, our method is much efficient,
as we do not need to wait that the system reaches the steady state.
Furthermore, the method allows us to reduce the fitting parameters
because the jamming transition point $\varphi_J$ is determined during
the procedure. Therefore, we believe that the AQC facilitates the
investigation of the criticality below jamming for frictionless
spherical particles, and hopefully for other models such as
non-spherical particles and frictional particles. For frictional
particles, it is reported that the waiting time and its distribution
show the power-law behaviors near the (shear) jamming transition
point~\cite{pastore2011flow,srivastava2019}. It is an interesting future
work to see if such critical behaviors of the dynamical quantities
appear under the AQC protocol.

The mean-field theory predicts that the correlated volume has the
fractal dimension $d_f=2$, irrespective of the spatial dimension
$d$~\cite{yan2016variational,during2016}. If this is the case,
the similar argument above Eq.~(\ref{005239_8Aug20}) leads to
\begin{align}
 \gamma = \frac{d-2}{d},\label{075855_24Dec20}
\end{align}
and that of Eq.~(\ref{073355_24Dec20}) leads to
\begin{align}
 \zeta = \frac{2d-1}{d-1}.\label{075900_24Dec20}
\end{align}
We obtained numerical results consistent with
Eqs.~(\ref{075855_24Dec20}) and (\ref{075900_24Dec20}) in three
dimension $d=3$. It is tempting to test if Eqs.~(\ref{075855_24Dec20})
and (\ref{075900_24Dec20}) hold in other $d$. In particular, $\gamma=0$
in $d=2$, suggesting that the participation ratio $P$ at $\varphi_J$ may
exhibit the logarithmic dependence on $N$, instead of a power law. This
deserves further study.

\begin{acknowledgments}
We warmly thank A.~Ikeda for discussions related to this work. We also
thank S.~Teitel and anonymous referees for useful comments. This project
has received funding from the JSPS KAKENHI Grant Number JP20J00289.
\end{acknowledgments}

\appendix

\section{$\varepsilon$ dependence}
\label{053220_24Dec20}

\begin{figure}[t]
\begin{center}
\includegraphics[width=8cm]{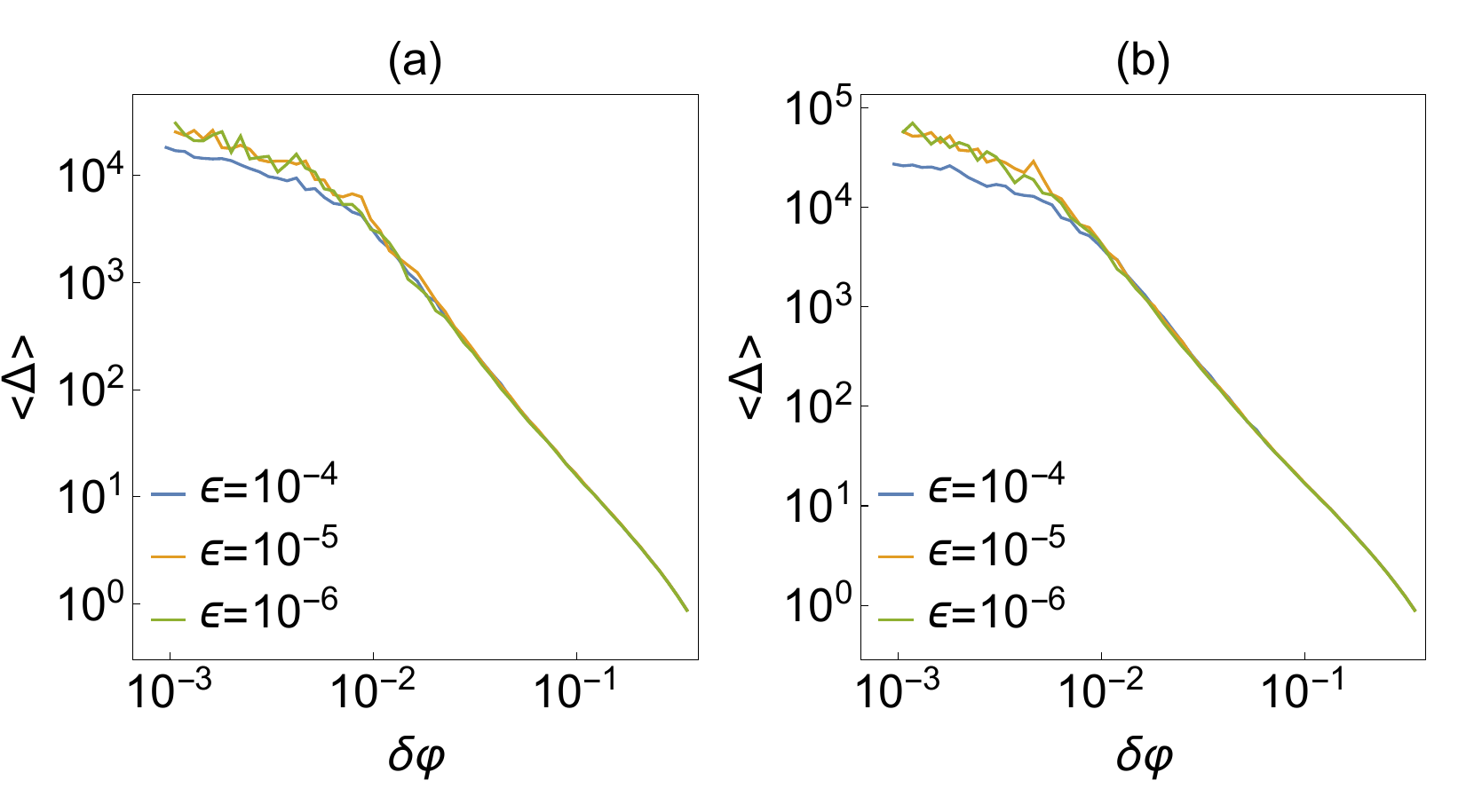} \caption{$\delta\varphi$
dependence of $\ave{\Delta}$ with varying $\varepsilon$ for $N=512$ (a)
and $N=1024$ (b).}  \label{164230_23Dec20}
\end{center}
\end{figure}

Here we show numerical results of examining the $\epsilon$ dependence in
our simulation protocol. Fig.~\ref{164230_23Dec20} presents the
$\varepsilon$ dependence of $\ave{\Delta}$ for $N=512$ and $N=1024$.
One can see that the results do not depend on $\varepsilon$ for
$\varepsilon\leq 10^{-5}$. The similar $\varepsilon$ dependencies of the
physical quantities have been previously reported by the numerical study
of the quasi-static shear~\cite{PhysRevE.83.031307}.  Therefore, in this
study, our simulation was performed with $\epsilon=10^{-5}$.

\bibliography{apssamp}

\end{document}